\documentclass[prd,aps,floats,preprintnumbers,preprint]{revtex4}

\usepackage{graphicx, epsfig}
 
\newcommand{\be}{\begin{equation}}
\newcommand{\ee}{\end{equation}}
\newcommand{\bea}{\begin{eqnarray}}
\newcommand{\eea}{\end{eqnarray}}

\begin{document}

\setlength{\unitlength}{1mm}

\title{Approximate Consistency Condition from Running Spectral Index
  in Slow-Roll Inflationary Models}
 
\author{Daniel J.H. Chung and Antonio Enea Romano}
\affiliation{
Department of Physics, University of Wisconsin, Madison, WI 53706, USA
}
 
\begin{abstract}
Density perturbations generated from inflation almost always have a
spectral index $n_s$ which runs (varies with the wavelength).  We
explore a running spectral index scenario in which the scalar spectral
index runs from blue ($n_s >1$) on large length scales to red
($n_s<1$) on short length scales.  Specifically, we look for a
correlation between the length scale at which $n_s-1=0$ and the length
scale at which tensor to scalar ratio $\mathcal{T}/\mathcal{S}$
reaches a minimum for single field slow roll inflationary models.  By
computing the distribution of length scale ratios, we conclude that
there indeed is a new approximate consistency condition that is
characteristic of running spectral index scenarios that run from blue
to red.  Specifically, with strong running, we expect 96\% of the slow
roll models to have the two length scales to be within a factor of 2,
with the length scale at which the tensor to scalar ratio reaching a
minimum longer than the wavelength at which $n_s-1=0$.
\end{abstract}

 
\maketitle
 
\section{Introduction.}
\label{introduction}
It is currently widely accepted that inflationary cosmological
scenarios offer most promising explanations for the initial conditions
for structure formation in our universe.  Almost all inflationary
scenarios predict that the primordial density perturbation spectrum
deviates slightly from a power law and is dominated by scalar density
fluctuations.\footnote{Exceptions to this can be found in
\cite{Vallinotto:2003vf}.}  Typically, the scalar density perturbation
spectrum is parameterized as
\begin{equation}
P_S(k)= A_S^2 (\frac{k}{k_*})^{n_s(k)-1}
\end{equation}
where $n_s(k)\approx 1$ is a scale dependent function which is usually
called the running (scalar) spectral index (running refers to the
change in the spectral index as a function of wavenumber $k$).
Combining observations of CMB, galaxy surveys, and Ly$\alpha$ forest, there
have been claims for evidence of strongly running spectral index (see
e.g. \cite{Peiris:2003ff}), but at the moment, combined data set
favors no spectral index running (e.g. \cite{Seljak:2004xh}).
Nonetheless, a significant running of the spectral index is still a
debatable possibility that will be settled by future experiments.

According to the CMB and 2dFGRS galaxy survey
data~\cite{Bennett:2003bz,Percival:2001hw},
\begin{equation}
\label{WMAP}
\frac{dn_s}{d\ln k}(k=0.05 \mbox{Mpc}^{-1}) = -0.03^{+0.016}_{-0.018} \ .
\end{equation}
Furthermore, as pointed out by \cite{Peiris:2003ff}, within the
context of single field inflationary models, there is some indication
that the spectral index quantity $n_s -1$ runs from positive values
(blue) on long length scales to negative values (red) on short length
scales (positive to negative within about 5 e-folds).

It is well known \cite{Lidsey:1995np} that one robust check of slow
roll inflationary scenario is what is usually referred to as the
single field self-consistency condition
\begin{equation}
n_T(k) = -2 \epsilon
\end{equation}
where $n_T$ is the spectral index of tensor perturbation power
spectrum parameterized as $P_T \propto k^{n_T}$ and $\epsilon$ is the
slow roll parameter characterizing the tensor to scalar power spectrum
ratio.  In \cite{Chung:2003iu}, it was pointed out that if running
occurs from blue to red, then there may be another approximate
consistency condition that we may observationally aim at checking
regarding inflationary scenarios.  Namely, within the single field
slow roll scenario, there should be an approximate coincidence between
the length scale $k_1$ at which
\begin{equation}
n_s(k_1)-1=0
\label{eq:k1def}
\end{equation}
and the length scale $k_2$ at which the tensor to scalar ratio reaches
a minimum, i.e.
\begin{equation}
\epsilon'(k_2)=0.
\label{eq:k2def}
\end{equation}  
If true, this kind of new consistency condition would be important
because currently there is only a very limited number of observational
consistency checks that could support the picture of inflationary
origin of density perturbations.

The degree to which 
\begin{equation}
k_1 \approx k_2
\label{eq:maincoincidence}
\end{equation}
depends on a) how fast $n_s(k)$ runs and b) the slow roll parameter
$\epsilon$ (this is true at least in the leading slow roll order
approximation).  Hence, it is not clear how compelling the
expectation of this coincidence is.  For example, does
Eq.~(\ref{eq:maincoincidence}) occur only for 1 out of 10,000
slow roll models of inflation with running spectral index or does it
occur for 9000 out of 10,000 slow roll models of inflation?
Furthermore, because of the strongly running behavior of the spectral
index, it was not clear that the leading order presented in
\cite{Chung:2003iu} was robust.  Hence, in this paper, we quantify the
degree to which one would find the approximate consistency condition a
compelling consistency check of single field inflationary scenario.
Furthermore, we check that the qualitative expectation based on first
order slow roll expansion is also valid at second order slow roll
expansion.  

In the spirit of \cite{Kinney:2003uw}, we examine a grid of slow roll
models and ask how the distribution of $k_1/k_2$ changes as a function
of increasing the requirement of running spectral index.  Expressed in
terms of difference in the e-folds of inflation (expressed as $k_1/k_2
= \exp(\Delta N)$), we find $\Delta N$ has a mean of about $-0.34$
with a $2\sigma$ width of $0.5$ when a strong running condition is imposed
as when both $0.15 <n(k=0.002\mbox{Mpc}^{-1})-1<0.3$ and
$-0.3<n(k=1\mbox{Mpc}^{-1})<-0.15$ are satisfied.  Hence, slow roll
inflationary scenarios predict that there should be a coincidence of
$k_1$ and $k_2$ within a factor of two if the spectrum runs from blue
to red, and the mismatch is such that $k_1>k_2$.  Furthermore, since
the $2\sigma$ width of the distribution without strong running
condition is about $1.2$, we find quantitative evidence that if the
spectral index runs strongly, there is an increase in the coincidence
as expected.

There are two main caveats to our results.  As discussed in Section
\ref{sec:discconc}, because the B-mode polarization sensitive to
tensor perturbations is expected to peak on relatively large angular
scales where it is measurable with current forecasts
\cite{Knox:2002pe,Cooray:2005xr} while the minimum of tensor to scalar
ratio lies on shorter angular scales, measurements testing the
proposed approximate consistency condition may be difficult.  If
strong running is relevant to cosmology, ingenuity of scientists in
the future will hopefully overcome this obstacle.  The second is that
the grid of models that we choose are sampled with equal weight for
the initial conditions of the slow roll equation.  Although this is
commonly found in the literature
\cite{Peiris:2003ff,Kinney:2003uw,Dodelson:1997hr,Boyle:2005ug} and
does not seem to be an unreasonable way to sample the possible set of
models, this kind of probability measure does not have any
justification from first principles.  A similar ambiguity of measure
plagues most ``landscape'' or anthropic principle arguments
(e.g. \cite{Banks:2003es} ).

This paper is organized as follows.  In the next section, we
analytically compute the coincidence in terms of slow roll parameters.
Following that we present numerical results. Measurement prospects are
then discussed in Section \ref{sec:discconc}.  In the
appendix, we collect some useful slow roll formulas used in the
derivations in the paper.  Throughout this paper, we use the
convention $M_p=1/\sqrt{G_N}$.

\setlength{\unitlength}{1mm}

\section{Analytical approximation}

Let us see why there should be an approximate consistency condition
as described in the introduction.  In \cite{Chung:2003iu}, it was
shown that within the context of a single real scalar field $\phi$
slow roll inflationary model with a potential $V(\phi)$, one can write
\begin{equation}
\frac{n_s-1}{\sqrt{2 \epsilon_V}} + \sqrt{2 \epsilon_V} =\pm
\frac{\epsilon_V'(\phi)}{\epsilon_V(\phi)} 
\label{eq:oldwork}
\end{equation}
where $\epsilon_V \equiv \frac{M_p^2}{16\pi}(V'/V)^2$ is the usual
inflationary slow roll parameter in terms of the potential, the upper
(lower) sign is for $V'(\phi)>0$ ($V'(\phi)<0$).  Since the tensor to
scalar ratio is given by $P_g/P_{\mathcal{R}}=16 \epsilon_V$, the vanishing of the right
hand side of Eq.~(\ref{eq:oldwork}) corresponds to the scale at which
the tensor to scalar power reaches an extremum.  Now, the observation
that was made was that if the spectrum runs from blue to red, then
$\pm \epsilon_V'$ will be positive initially, but as $n_s-1$ becomes
negative to the point of canceling the $2 \epsilon_V$, $\epsilon_V'$
will vanish.  As long as $\epsilon_V$ is small and/or $n_s-1$ runs
strongly negative, the scale at which $n_s-1$ crosses zero will
coincide with the extremum of $\epsilon_V$ which in turn corresponds
to the extremum of the tensor to scalar ratio.

In this section we will extend the previous derivation to second order
in slow roll and confirm the qualitative validity of the previous
results.  This is reassuring given that the validity of leading order
results was questionable in light of strongly running behavior.  To
accomplish this task, we will work with Hubble flow slow roll
expansion techniques (see, e.g. \cite{Lidsey:1995np}).  The advantage
of the Hubble flow approach is that it is simpler to go to higher
orders in derivative expansion.

Using the conventions of \cite{liddle94}, we use the definitions for
the slow roll parameters in terms of the Hubble function
$H\left(\phi\right)$ as follows:
\begin{eqnarray}
\epsilon &\equiv& {M_p^2 \over 4 \pi} \left({H'(\phi) \over
 H(\phi)}\right)^2\cr
\eta\left(\phi\right) &\equiv& {M_p^2 \over 4 \pi} 
\left({H''\left(\phi\right)
\over H\left(\phi\right)}\right)\\
\sigma &\equiv& {M_p^2 \over \pi} \left[{1 \over 2} \left({H'' \over
 H}\right) -
\left({H' \over H}\right)^2\right]\cr
{}^\ell\lambda_{\rm H} &\equiv& \left({M_p^2 \over 4 \pi}\right)^\ell
{\left(H'\right)^{\ell-1} \over H^\ell} {d^{(\ell+1)} H \over d\phi^{(\ell +
1)}}.
\end{eqnarray}
Here, the Hubble function is defined by the Hamilton-Jacobi equations
\begin{eqnarray}
\dot{\phi} & = & -\frac{M_p^2}{4 \pi} \frac{dH}{d\phi} \\
\left(\frac{dH}{d\phi} \right)^2 - \frac{12 \pi}{M_p^2} H^2 & = &\frac{-32 \pi^2}{M_p^4} V(\phi)
\end{eqnarray}
which has the advantage of having a readily solvable model of power
law inflation.  As is well known, Hubble flow slow roll expansion that
we summarize below is a derivative expansion in $H(\phi)$ about the
power law inflationary model.

We used $N$ as the measure of time during inflation, the number of
e-folds before the end of inflation, which increases as one goes {\em
backward} in time :
\begin{equation}
{d \over d N} = {d \over d\ln a} = { M_p \over 2 \sqrt{\pi}}
\sqrt{\epsilon} {d \over d\phi},
\label{eq:ddn}
\end{equation}
with the sign convention
\begin{equation}
\sqrt{\epsilon} \equiv + {m_{\rm p} \over 2 \sqrt{\pi}} {H' \over H}.
\end{equation}
These equations implies a relationship of $\epsilon$ to $H$ and $N$:
\begin{equation}
\label{eqepsilonfromN}
{1 \over H} {d H \over d N} = \epsilon.
\end{equation}

The evolution of the higher order parameters during inflation is
determined by a set of ``flow'' equations \cite{hoffman00,schwarz01,kinney02},
\begin{eqnarray}
{d \epsilon \over d N} &=& \epsilon \left(\sigma + 2
\epsilon\right),\cr {d \sigma \over d N} &=& - 5 \epsilon \sigma - 12
\epsilon^2 + 2 \left({}^2\lambda_{\rm H}\right),\cr {d
\left({}^\ell\lambda_{\rm H}\right) \over d N} &=& \left[
\frac{\ell - 1}{2} \sigma + \left(\ell - 2\right) \epsilon\right]
\left({}^\ell\lambda_{\rm H}\right) + {}^{\ell+1}\lambda_{\rm
H}.\label{eqfullflowequations}
\end{eqnarray}
As one can see, in general any slow roll parameter could be expressed
in terms of one single slow roll and its derivatives.  Some of these
relationships are explicitly given in the appendix. Note that
$\epsilon$ has the interpretation through the fact that tensor to
scalar power spectrum is given by 
\begin{equation}
\frac{P_g}{P_{\mathcal{R}}}=16 \epsilon
\end{equation}
where $P_g$ and $P_{\mathcal{R}}$ are the usual power spectrum for tensor and
scalar gauge invariant metric perturbations.

Let us now solve for the difference between the length scale at which
$\epsilon$ reaches an extremum ($1/k_2$) and the length scale at which
$n_s-1$ crosses 0 ($1/k_1$).  Instead of giving the length scale in
terms of the inverse wave vector, we will express it terms of the
difference in the horizon exit e-folds $\Delta N$.  To translate that
into ratio of lengths, one simply has 
\begin{equation}
\ln(\frac{k_2}{k_1})=\Delta N=N_1-N_2 + \mathcal{O}(\ln(H_2/H_1))
\label{eq:deltan}
\end{equation}
where $H_2$ is the expansion rate when $k_2$ left the horizon and
$H_1$ is the expansion rate when $k_1$ left the horizon.  Since
$\ln(H_2/H_1)$ is of order of the slow roll parameter during the early
stages of inflation, one can disregard this term when interpreting
$\Delta N\equiv N_1-N_2$ as long as it is larger than about $0.1$.
Defining $N_{0}$ as the point where the extremum of $\epsilon$ occurs,
i.e. $\frac{d\epsilon}{dN}|_{N_{0}}=0$, we can linearize $n_s-1$ around
it to solve approximately for $\Delta N$.  Using the formulas given in
the appendix, we can write
\begin{equation}
n_s(N_{0}+\Delta N) -1 \simeq n_s(N_{0})-1+\frac{dn_s}{dN}|_{N_{0}}\Delta N
\end{equation}
\begin{equation}
\frac{dn_s}{dN}=-\frac{2\epsilon^2\dot{\epsilon}-\dot{\epsilon}^2-\epsilon\ddot{\epsilon}}{\epsilon^2}
\end{equation}
\begin{equation}
n_s-1=\frac{1} {4\epsilon^2}\{ (-3+C)\dot{\epsilon}^{2}+ \left[4\epsilon+2\epsilon^2(C+3)\right]\dot{\epsilon} +\ddot{\epsilon}(3-C)-8\epsilon^3(\epsilon+1)\}.
\end{equation}
Solving $n_s(N_{0}+\Delta N)-1=0 $ for $\Delta N$, we obtain
\begin{equation}
\Delta N_{analytical}\approx \frac {C-3}{4}+\frac
       {2\epsilon(1+\epsilon)}{\frac{dn_s}{dN}(N_{0})}.
\end{equation}
The constant $C \equiv 4 (\ln{2} + \gamma) - 5 = 0.0814514$, where
$\gamma \simeq 0.577$ is Euler's constant.
As it can be seen from this formula $\Delta N$ clusters around the
negative value $-0.67$ and it can be positive or negative.  We can
also observe that stronger running (associated with a higher slope for
$n_s$) drifts $\Delta N$ toward more negative values, and that this effect is modulated by the value of $\epsilon$ at $N_{0}$.

\begin{figure}
	\centering
		\includegraphics{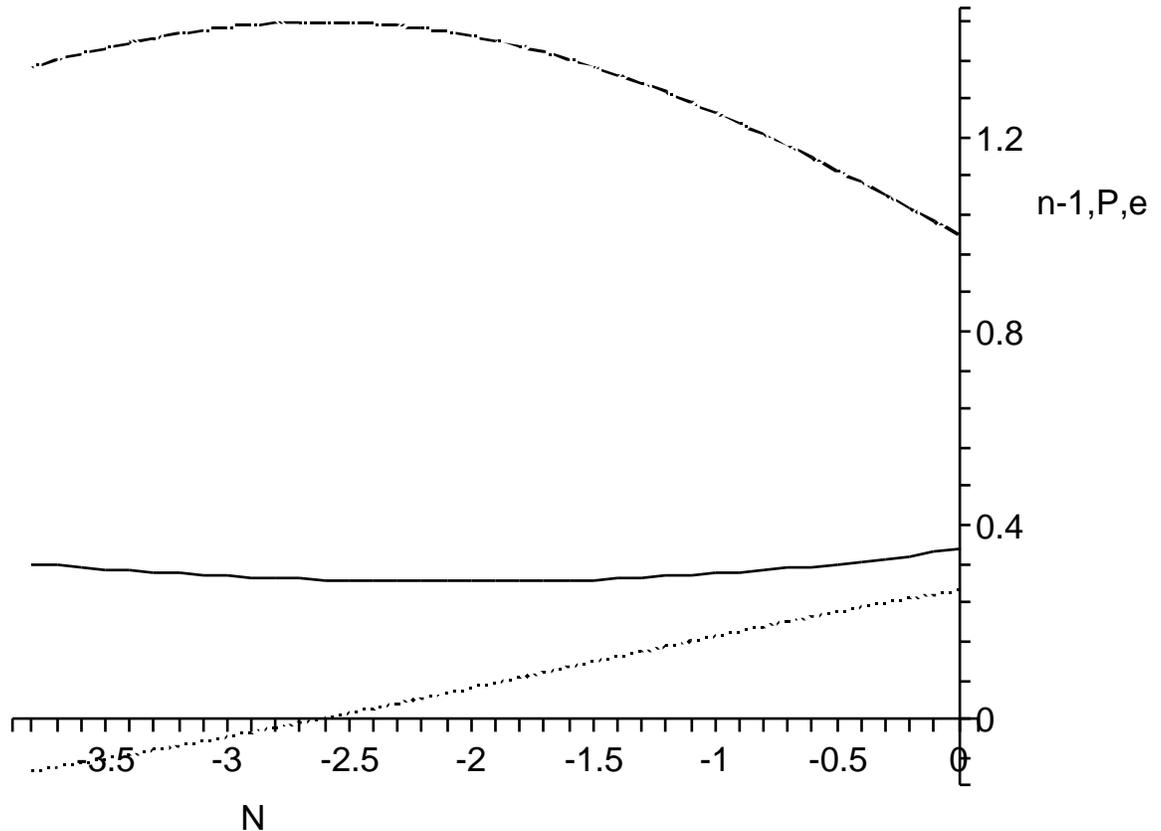}
	\caption{From the graph of $\epsilon$ , $P_{\mathcal{R}}$
	(scalar power spectrum), and $n_s-1$ as functions of the
	number of e-folds $N$, it can be seen that the extrema of
	$\epsilon$ and $P_{\mathcal{R}}$ are closed to the point where
	$n_s-1$ is zero.  The dot dashed curve is
	$P_{\mathcal{R}}/P_{\mathcal{R}}(0)$, the solid curve is
	$50\epsilon$, and the dotted curve is $n_s-1$.  The horizontal
	axis is oriented such that inflation ends at $N<-3.5$
	(i.e. time flows right to left).}
	\label{fig:P-e-n-plot}
\end{figure}

We will see that numerical results agree well with the analytical
approximation.  Before going onto numerical analysis, we would like to
note that beyond Eq.~(\ref{eq:oldwork}), there may be many more
``approximate consistency'' conditions associated with strong running
since we are merely labeling various features of the slow roll
equations.  For example, as it can be inferred already from the first
order formula $ P_k^{\zeta} = \frac{8 V_k}{3 \epsilon M_p^4} $ the
power spectrum will have an extremum roughly coinciding with the
extrememum of $\epsilon$ .  To check this numerically, we can examine
an example of an inflationary model (as explained in the next
section).  As it can be seen in Fig.~\ref{fig:P-e-n-plot}, the scalar power
spectrum has an extremum very closed to the point where $n_s-1$ is zero.
We could proceed similarly to the way we have done for
Eq.~(\ref{eq:maincoincidence}), comparing analytical approximation to
numerical data for this ``approximate consistency'' condition, which
in this case would not depend on the strength of the running of the
spectral index, but could be another test for more general inflation
models.

\section{Numerical approach}
In this section, we study our proposed approximate consistency
condition question numerically.  One scientific question we would like
to address is, ``What is the expected coincidence of the value $k_1$
satisfying $n_s(k_1)-1=0$ with the value $k_2$ satisfying
$\epsilon'(\phi_{k_2})=0$?'' The second question is whether a stronger
negative running of the spectral index makes the expected $|k_1-k_2|$
significantly smaller.

To this end, we examine a particular grid (to be specified below) of
slow roll inflationary models.  In this grid of models, we make a set
of ``cuts'' which select out subsets of the models with scalar
spectral index running from blue to red.  In this way, we can make a
histogram of models as a function $\Delta N$ (defined in
Eq.~(\ref{eq:deltan})).  The peak and the width of the histogram
quantifies the expected coincidence of $k_1$ with $k_2$, answering our first
question.  We then make the cuts more stringent such that only those
slow roll models with more strongly running spectral indices are
included in the histogram.  The degree to which the width of the
histogram changes as we make the required running stronger answers our
second question.

Let us now explain the details which closely follows the methods of
\cite{Kinney:2003uw}.  Assigning a set of initial conditions to
Eqs.~(\ref{eqfullflowequations}) and integrating them specifies a flow
of the slow roll function.  Each of the initial conditions corresponds
to a particular choice of inflaton potential and the inflaton initial
conditions.  We choose initial values for the parameters at random
from the following ranges, assuming a uniform probability distribution
:
\begin{eqnarray}
N &=& [40,70]\cr
\epsilon &=& \left[6\ 10^{-76},0.6\right]\cr
\sigma &=& \left[-0.5,0.5\right]\cr
{}^2\lambda_{\rm H} &=& \left[-0.05,0.05\right]\cr
{}^3\lambda_{\rm H} &=& \left[-0.025,0.025\right],\cr
&\cdots&\cr
{}^{M+1}\lambda_{\rm H} &=& 0.\label{eqinitialconditions}
\end{eqnarray}
We truncated the expansion to order 5 by setting ${}^{6}\lambda_{\rm
H} = 0$.  We calculate the values of the tensor/scalar ratio $r$, the
spectral index $n$, and the ``running'' of the spectral index $d n /
d\ln k$ according to: \cite{liddle94,stewart93}
\begin{equation}
r   = 10 \epsilon \left[1 - C \left(\sigma + 2
 \epsilon\right)\right],\label{eqrsecondorder}
\end{equation}
\begin{equation}
n_s - 1 = \sigma - \left(5 - 3 C\right) \epsilon^2 - {1 \over 4} \left(3
- 5 C\right) \sigma \epsilon + {1 \over 2}\left(3 - C\right)
\left({}^2\lambda_{\rm H}\right)\label{eqnsecondorder}
\end{equation}
Note that the variable $r$ here is defined with a different
normalization than in \cite{Peiris:2003ff} where $r\approx 16
\epsilon$ to leading slow-roll order.  A comoving scale $k$ crossed
the horizon a number of e-folds $N(k)$ before the end of inflation :
\cite{lidsey97}
\begin{equation}
N(k) = 62 - \ln \frac{k}{a_0 H_0} - \ln \frac{10^{16}
        {\rm GeV}}{V_k^{1/4}}
        + \ln \frac{V_k^{1/4}} {V_e^{1/4}} - \frac{1}{3} \ln
        \frac{{V_e}^{1/4}}{\rho_{{\rm RH}}^{1/4}} \, .
\label{eq:comovingnk}
\end{equation}
Here $V_k$ is the potential at horizon exit, $V_e$ is the potential at
the end of inflation, and $\rho_{{\rm RH}}$ is the energy density
after reheating.  Note that since slow roll inflation evolves toward
decreasing potential, we can write
\begin{equation}
V_e = f_k^4 V_k
\label{eq:vevkrel}
\end{equation}  
where $f_k < 1$ is a function of $k$ that varies as to keep $V_e$
independent of $k$.   Furthermore, since the reheating energy density
must be smaller than $V_e$, we can write
\begin{equation}
\rho_{{\rm RH}}= \gamma^4 V_e
\label{eq:verhorel}
\end{equation}
where the proportionality constant satisfies $\gamma<1$.  Using
Eq.~(\ref{eq:comovingnk}), derivatives with respect to wavenumber $k$
can be expressed in terms of derivatives with respect to $N$ as
\cite{liddle95}
\begin{equation}
{d \over d N} = - \left(1 - \epsilon\right) {d \over d \ln k},
\end{equation}
Hence, the running of the spectral index can be expressed to third
order in slow roll parameters as 
\begin{eqnarray}
\frac{dn_s}{d \ln k} &= &\frac{1}{4 (1-\epsilon)} 
\left[ 
48\,{{\epsilon}}^{2}+ \left( 44+12\,C \right) {{\epsilon}}^{3}+
\left( -9\,C+31 \right) {{\epsilon}}^{2}\sigma+ \left( -5\,C+3 \right) {\epsilon}\,{\sigma}^{2}
\right. \nonumber
\\
& & 
\left.
+
 \left( -10\,C+6
 \right) ({}^{2}\lambda_{\rm H}) {\epsilon}+ \left( C-3
 \right) ({}^{2}\lambda_{\rm H})\sigma  + 
+ 20\,{\epsilon}\,\sigma-8\,({}^{2}\lambda_{\rm H})+ \left( 2\,C-6
  \right) ({}^{3}\lambda_{\rm H})
\right]
\end{eqnarray}

We will now use superscripts a and b to denote quantities evaluated at
or corresponding to the length scale $k^a=0.002 \mbox{ Mpc}^{-1}$ or
$k^b=1 \mbox{ Mpc}^{-1}$.  The algorithm for generating the histogram
of Fig.~1 can then be described as follows:
\begin{enumerate}
\item A set of random initial conditions was generated for the value
of the slow roll parameters at $N^{a}$.  The following constraints are imposed:
  \begin{equation}r^{a}<0.5 \label{eq:rcond}\end{equation}
  $$n^{a}_{l}<n_s^{a}-1<n^{a}_{u}$$
  $${ \frac{dn_s}{d \ln k} }^{a} < 0 $$
where $n^{a}_{u}$ is fixed to 0.3 and $n^{a}_{l}$ varies from 0 to 0.225 for different cuts with a 0.075 increment.
If these constraints are respected, proceed to step 2 otherwise go
back to step 1.  

\item Integrate the flow equation (using LSODA from
ODEPACK \cite{Hindmarsh:1983}) to $N^{b}$ and then check the
following constraints:
$$n^{b}_{l}<n_s^{b}-1<n^{b}_{u}$$
where $n^{b}_{l}$ is fixed to -0.3 and $n^{b}_{u}$ varies from 0 to
-0.225 for different cuts with a 0.075 increment. 
If these are respected go to step 3 otherwise go to step 1.

\item Evolve forward in time ($d N < 0$) until inflation ends
($\epsilon > 1$).If the total number of e-folds $N$ from the beginning
to the end of inflation is in the range [40,70] add this model to the
ensemble of acceptable models. If not go back to step 1.

\item Repeat step 1 through 3 until the desired number of acceptable
models have been found.
\end{enumerate}

Once the set of acceptable models has been generated we solve
numerically the following two equations:
\begin{equation}
{d \epsilon \over d N}|_{N_2}=0
\label{eq:epsprime}
\end{equation}
\begin{equation}
n_s|_{N_1}-1=0
\end{equation} It is important to observe that the presence of the
square root of $\epsilon$ in the definition of the differential
operator $d \over d N$ (Eq.~(\ref{eq:ddn})) causes numerical solutions
to Eq.~(\ref{eq:epsprime}) to be difficult for small values of
$\epsilon$, since small values of the derivative may be associated to
the smallness of $\epsilon$ more than to the actual presence of an
extremum.  In other words, even though the derivatives are taken with
respect to $N$, for inflaton potentials with very small $\epsilon$,
the derivatives become very small, requiring high numerical precision.
For this reason and to discard inflection points, extrema are
identified numerically using the second derivative information as
well.

Before making the histogram, we must further check that the models in
the histogram are reasonable from an inflationary phenomenology point
of view.  This is done in lieu of executing precision fits to all
available data since it has been well demonstrated by previous
analyses (see for example \cite{Kinney:2003uw}) that the current CMB
data are not extremely constraining in terms of the details of the
slow roll inflationary models.  Hence, our histogram should not be
sensitive to the lack of precision in our cuts.  On the other hand,
given that the combined fits including SDSS data as carried out by
\cite{Seljak:2004xh} prefer small running, the strongly running cases
of the current analysis may have been statistically disfavored had
precision fits to combined data been made.  Given that most of the
numerically generated models do not have strongly running spectral
index for the weakest cut, the results of the weakest cut should be
robust with respect to imposing additional fit constraints, while the
results of the stronger cuts should be interpreted with appropriate
caution.

Let us now write explicitly the constraint equations.  Since the WMAP
analysis~\cite{Peiris:2003ff} gives
\begin{equation}
P^\zeta (k_a = 0.002 \mbox{Mpc}^{-1}) =  ( 2.95\times 10^{-9})(0.75
\pm 0.09) \ ,
\label{eq:powerobserve}
\end{equation}
we use the approximate formula for the power spectrum
\begin{equation}
P_k^{\zeta} = \frac{8 V_k}{3 \epsilon M_p^4}
\label{eq:approxpzeta}
\end{equation}
to impose a constraint on $\epsilon$ and $V_k$.
The requirements of having sufficient energy density at the end of
inflation to reheat the universe to a temperature consistent with big
bang nucleosynthesis gives
\begin{equation}
(10^{-2} \mbox{GeV})^4  < \rho_{\rm{RH}} < V_e <V_k. \label{eq:enoughreheat} 
\end{equation}
We also know from observations that $r_a <\mathcal{O}(1)$ and that
Eq.~(\ref{eq:powerobserve}) is true.  
Substituting $V_k$ obtained from Eqs.~(\ref{eq:comovingnk}),
(\ref{eq:vevkrel}), and (\ref{eq:verhorel}) into
Eq.~(\ref{eq:approxpzeta}) and solving for $\epsilon$ we obtain
\begin{eqnarray}
 6\times 10^{-76}< \epsilon_a& <&  0.05
\label{eq:yconst}\\
\epsilon_a & = & e^{4(N(k_a)-61.8)} y(k_a) 
\label{eq:epsyrel}
\end{eqnarray}
where $y(k_a)\equiv f_{k_a}^4\gamma^{-4/3} $ and the upper bound comes
from Eq.~(\ref{eq:rcond}) and $r=10 \epsilon$ (again note that the
normalization used in this paper is given by
Eq.~(\ref{eqrsecondorder}), and specifically is not $r=16
\epsilon$).
Consider how Eq.~(\ref{eq:epsyrel}) gives a constraint.  Note that for
any given flow trajectory, because of Eqs.~(\ref{eq:powerobserve}) and
(\ref{eq:approxpzeta}), $V_{k_a}$ is fixed.  Furthermore, the flow
trajectory itself determines $f_{k_a}$ (the value of the potential at
the end of inflation).  Hence, the only adjustable parameter is
$\gamma$ which determines the reheating temperature.  For every flow
trajectory with $N(k_a)$ fixed, Eq.~(\ref{eq:epsyrel}) then provides
constraints on the initial conditions of inflation.  For example,
consider a flow trajectory having $N(k_a)=70$.  Eq.~(\ref{eq:epsyrel})
gives
\begin{equation}
f_{k_a}^4  < 0.05 \exp( -4 [N(k_a)- 61.8])\sim 10^{-16}
\label{eq:fkconstraint}
\end{equation}
This is a significant constraint since this says that the energy
density at the end of inflation is at least $10^{-16}$ smaller than
the the energy density at $70$ e-foldings before the end of
inflation.  The physical reason for Eq.~(\ref{eq:yconst}) is that
upper and lower bounds on the potential (coming from tensor
perturbations and reheating temperature, respectively) directly
translates into upper and lower bounds $\epsilon$ because of
Eqs.~(\ref{eq:powerobserve}) and (\ref{eq:approxpzeta}).  The physical
origin of Eq.~(\ref{eq:epsyrel}) is that any mismatch in the
inflationary stretching of the wavelength and physical $k/a=0.002
\mbox{Mpc}^{-1}$ must be compensated by postinflationary expansion
which is related to the potential energy at the end of inflation which
in turn is related to $\epsilon$ through Eqs.~(\ref{eq:powerobserve}),
(\ref{eq:approxpzeta}), and (\ref{eq:vevkrel}).  In practice,
Eq.~(\ref{eq:fkconstraint}) does not play a significant role
in the numerical exploration since the number of e-foldings is
generically small when there is significant running.


\begin{figure}
\psfig{figure=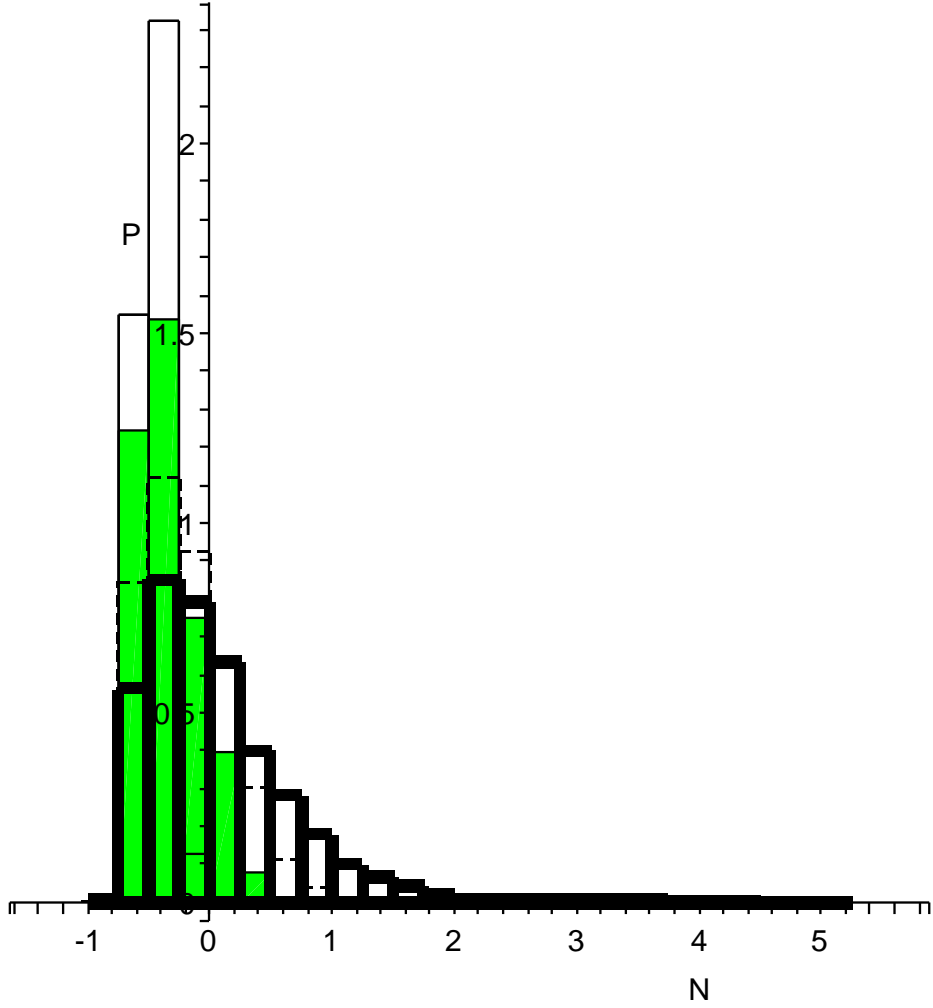,width=6.0in}
\vskip .1 in
\caption{Frequency histogram (hist.) for the difference between
solutions for different running of the spectral index, i.e. different
$n_a$ (spectral index at wave vector $k^a=0.002$Mpc$^{-1}$) and $n_b$
(spectral index at wave vector $k^b=1$Mpc$^{-1}$).  Thick solid
hist.~gives the results for \{ $0<n_a-1<0.3$ , $-0.3<n_b-1<0$\};
dashed hist.~for \{ $0.075<n_a-1<0.3$, $-0.3<n_b-1<-0.075$\}; the
green hist.~for \{$0.15<n_a-1<0.3$ , $-0.3<n_b-1<-0.15$\}; the
solid hist.~for \{$0.225<n_a-1<0.3$ , $-0.3<n_b-1<-0.225$\}. The
area under each histogram is normalized to unit area.}
\label{figpk}
\end{figure}

Our final histogram can be seen in Fig.~\ref{figpk} and its
distributional characterization can be seen in Table
\ref{tbl:mainresult}.  To have a measure of how $\Delta N$ clusters
around zero we followed this procedure: a) we have divided $\Delta N$
in the range $[-1,5.25]$ into 25 intervals of width 0.25 each
containing $n_i$ models and then calculated for each of these
intervals the average $\langle{\Delta N}\rangle_{i}$ for the models in
that interval and the standard deviation $\sigma_{\Delta
N_{i}}=\frac{1}{\sqrt{n_i}}$ ; b)we calculated
$m_2=\Sigma_{i=1}^{25}\frac{\langle{\Delta
N}\rangle^2_{i}}{\sigma^2_{\Delta N_{i}}}$, a weighted second order
moment around 0 of the 25 averages of $\langle{\Delta N}\rangle_{i}$
for the intervals defined above, using the inverse of the
corresponding variance $\sigma_{\Delta N_{i}}$ as weight.  We also
similarly calculated the weighted standard deviation $\sigma_{m_2}$.
The total number of slow roll models we considered to construct the
histogram was around $10^7$, and the number of models fulfilling the
cuts is shown in the $N_{md}$ column of Tbl.~\ref{tbl:mainresult}.
  

\begin{table}
\begin{tabular}{|r|r|r|r|r|r|r|r|}
\hline $k=0.002 Mpc^{-1}$ & $k=1 Mpc^{-1}$ & $m_2$ & $\sigma_{m_2}$ &
$\langle \Delta N \rangle$ & $\sigma_{\Delta N}$ & $P( \langle \Delta N \rangle \pm 2\sigma_{\Delta N} )$ & $N_{md}$\\ \hline 
$0<n_s-1<0.3$ & $-0.3<n_s-1<0$ & 0.644 & 0.022 & -0.073 & 0.643 & 0.961 & 2366        \\ \hline 
$0.075<n_s-1<0.3$ & $-0.3<n_s-1<-0.075$ & 0.395 & 0.023 & -0.184 & 0.356 & 0.964 & 880\\ \hline 
$0.15<n_s-1<0.3$ & $-0.3<n_s-1<-0.15$ & 0.416 & 0.032 & -0.338 & 0.25 & 0.956 & 203   \\ \hline
$0.225<n_s-1<0.3$ & $-0.3<n_s-1<-0.225$ & 0.462 & 0.036 & -0.450 & 0.12 & 0.968 & 31  \\ \hline
\end{tabular}
\caption{Basic distributional characterization of Fig.~\ref{figpk}.
$m_2=\sqrt{\langle (\Delta N)^2\rangle}$ characterizes the deviation
from 0.  $P (\langle \Delta N \rangle \pm 2\sigma_{\Delta N} )$ is the
fraction of the models within 2$\sigma_{\Delta N}$ of $\langle \Delta
N\rangle$. $N_{md}$ is the number of models analyzed for the given range of the $n_s$.}
\label{tbl:mainresult}
\end{table}

Notice in Table \ref{tbl:mainresult} that both $|\langle \Delta N
\rangle \pm \sigma_{\Delta N} |<1$ and $|m_2 \pm \sigma_{m_2}|<1$
regardless of the strength of running.  The column labeled `` $P(
\langle \Delta N \rangle\ \pm 2\sigma_{\Delta N} )$'' quantifies (at
least with respect to our choice of model sampling) the extent to
which we should expect a coincidence between $k_1$ and $k_2$: i.e. we
should expect with roughly ``95\% confidence'' that the length scale
for which the tensor to scalar ratio reaches a minimum will coincide
with the length scale for which the $n_s-1=0$ up to a factor of about
2 and $k_1>k_2$ (i.e. wavelength at which the tensor to scalar ratio
reaches a minimum is longer than the wavelength at which $n_s-1$ goes
through a zero).  As far as the effect of different strengths of
running is concerned, one sees in Table \ref{tbl:mainresult} that
stronger running corresponds to
a larger magnitude of $|\langle \Delta N \rangle|$ and a more negative
value of $\langle \Delta N \rangle$.  Although ``statistically''
marginal, the smaller $\sigma_{\Delta N}$ means that strong
running seems to narrow the distribution of $\Delta N$.

\section{Discussion and Conclusions \label{sec:discconc}}
In this paper, we have studied the robustness of the approximate
condition suggested by \cite{Chung:2003iu}.  We have quantified
distributionally the extent to which one should consider the
approximate consistency condition to be a prediction of single field
slow roll inflation.  We find that if the spectrum runs from blue to
red, we should look for an approximate coincidence between the wave
vector $k_1$ satisfying $n_s(k_1)-1=0$ and the wave vector $k_2$
satisfying $\epsilon'(k_2)=0$, up to a factor of about 2.  According
to a WMAP analysis \cite{Peiris:2003ff}, this should occur at $k/a_0
\sim 0.02 \mbox{Mpc}^{-1}$.

Given that the consistency condition is only approximate, even if one
finds that observations contradict the consistency condition (for
example, if observations deduce $k_1/k_2 <1$) it will be very
difficult to make any rigorous conclusions about the models of
inflation.  Furthermore, given that we have sampled each of the
initial condition ranges with equal weight, it is not clear what the
distribution that we computed has to do with the real world.  (A
similar argument can be made about much of the current literature that
use similar sampling
\cite{Peiris:2003ff,Kinney:2003uw,Dodelson:1997hr,Boyle:2005ug}.)
Nonetheless, if this set of initial sampling turns out to be a good
approximation to the real inflationary model distributions coming from
a more fundamental theory, tests coming from this kind of approximate
consistency condition may be useful.  To turn this around, there may
be a way to classify the vacua of more fundamental theories according
to whether they satisfy roughly equal probability sampling of the slow
roll initial condition or whether they prefer a very specific
distribution which makes the approximate consistency condition more
(or less) stringent.

Before concluding, let us briefly consider the measurement prospects
for the approximate consistency condition.  According to current ideas
of anticipated measurements
\cite{Kamionkowski:1996ks,Kamionkowski:1996zd,Zaldarriaga:1996xe,Seljak:1996ti}, B-mode polarization observations are necessary to reliably infer
the tensor perturbation amplitudes.  As is well known, B-mode
polarization has the advantage over the E-mode polarization in that at
the last scattering surface, the B-mode polarization cannot be
generated by scalar perturbations alone because to leading order, the
scalar perturbations do not generate nonzero $\{l=2,m =\pm 2\}$
amplitudes in the temperature anisotropy while the tensor modes can.
Hence, in principle, assuming no large contamination from vector
perturbations, B-mode polarization measurement offers a direct way to
infer tensor perturbation amplitudes.

The difficulty with measuring the B-mode polarization spectrum coming
from gravity waves is that because its amplitude is proportional to
the local quadrupole anisotropy induced by the time derivative
$\dot{h}$ (where $h$ is the amplitude of the gravitational wave)
localized at the last scattering surface (due to the requirement of
Thomson scattering), its amplitude is suppressed at least by order
$k/(a_R H_R)$ on long wavelengths with respect to the temperature
anisotropy (where $H_R$ and $a_R$ are the Hubble radius and the scale
factor at recombination, respectively).  Furthermore, since
gravitational wave mode amplitudes inside the horizon are diluted by
the expansion of the spatial volume, the transfer function should
behave approximately as $\frac{1}{1+(\frac{k}{a_0})^2(\frac{a_R}{a_0})
\frac{1}{H_0^2}}$ where $a_0$ is the scale factor today and $a_R/a_0
\sim 10^3$ is the redshift to the last scattering surface.  (Here, we
have considered only the modes that entered the horizon after matter
domination, i.e. $k/a_0< H_0 \sqrt{a_0/a_{eq}}$.)  Hence, the
B-polarization temperature perturbation amplitude can be written as
\begin{equation}
k^{3/2}\Theta_{B,l}(k) \sim 10^{-1} \frac{k}{a_R H_R} \frac{\sqrt{P_h(k)}}{1+(\frac{k}{a_0})^2(\frac{a_*}{a_0}) \frac{1}{H_0^2}} j_l ( 3.5k/(a_0 H_0))
\end{equation}
where $P_h$ is the power spectrum of the tensor perturbations.  (For
discussions of analytic treatments of CMB polarization, see
e.g. \cite{Hu:1997hp,Pritchard:2004qp,Hu:1997mn,seljaknotes,dodelsonbook}.)
This peaks at $k/a_0 =H_0 \sqrt{a_0/a_*} \sim 1/(100 \mbox{Mpc}) $,
and even there, the amplitude is suppressed by about $10^{-2}$
relative to the temperature anisotropies if the tensor perturbation
amplitude is about the same as scalar perturbations amplitudes.

To make the situation slightly worse, $k_1$ is expected to be at
around $k_1/a_0 \sim 1/(50 \mbox{Mpc})$. (Note that to map the wave
vectors $k/a_0$ to the multipole moment number $l$, one can use the
approximate formula $\frac{k}{a_0} \sim \frac{l}{3.5} H_0 \approx
\frac{l}{15,000 \mbox{ Mpc}}$.)  This means that to actually measure
$k_1/k_2$, one must go to short scales ($l\sim 300$) while the
B-mode polarization measurements prefer large length scales ($l \sim
90$; for more accurate plots of examples, see
e.g. \cite{Knox:2002pe,Cooray:2005xr}).  Hence, experimental
confirmation of the approximate consistency conditions will be a
challenge.

There is one more obstacle that makes it difficult for tensor
perturbation amplitudes to be extracted from B-mode polarization
measurements.  This is due to the fact that E-modes can be converted
into B-modes through gravitational lensing
\cite{Zaldarriaga:1998ar,Seljak:2003pn}.  Hence, obtaining a tensor
spectrum to check the approximate consistency condition requires
success in extracting the tensor perturbation contribution even at
relatively short length scales, around where the lensing contribution
most likely dominates over the gravitational wave signal.  This
contamination from gravitational lensing can be subtracted out, at
least in principle, if the lens distribution can be accurately
deduced.

Despite the difficulties, we are optimistic that the ingenuity of
researchers in the future will allow a test of the approximate
consistency conditions considered here.

%
%

\section{Acknowledgment}
We thank Niayesh Afshordi and Scott Dodelson for useful discussions
and helpful, detailed comments on the manuscript.

\appendix
\section{Slow Roll Formulae}

Here we report some formulas which can be used to express slow roll
parameters in terms of $\epsilon $ and its derivatives respect to the
field $\phi$ or the operator $d \over d N$.  Denoting $\frac{d
\epsilon}{d \phi}=\epsilon^{'}$ and $\frac{d \epsilon}{d
N}=\dot\epsilon$ :
\begin{eqnarray}
  \eta & = & {\frac {4\,{{\epsilon}}^{3/2}\pi +{\epsilon^{'}}\,\sqrt {\pi }m}{4\pi
\,\sqrt {{\epsilon}}}} \cr
	\sigma &=& {\frac {-4\,{{\epsilon}}^{3/2}\pi +{\epsilon^{'}}\,\sqrt {\pi }M_p}{2\pi \,\sqrt {{\epsilon}}}} \cr
  {}^2\lambda_{\rm H}  &=&	{\frac {16\,\pi \,{{\epsilon}}^{3}+12\,{{\epsilon}}^{3/2}
	\sqrt {\pi }{\epsilon^{'}}\,M_p-{{\epsilon^{'}}}^{2}{M_p}^{2}+2\,{\epsilon^{''}}\,{M_p}^{2}{\epsilon}}{{16\pi\epsilon}\ }} \cr
  \eta & = & {\frac {\pi \, \left( 2\,{{\epsilon}}^{2}+{\dot\epsilon} \right) }{
  2\pi\,{\epsilon}}} \cr
  \sigma &=&\frac{\dot\epsilon-2\epsilon^2}{\epsilon} \cr
  {}^2\lambda_{\rm H} &=&{\frac {2\,{{\epsilon}}^{4}+3\,{{\epsilon}}^{2}{ \dot\epsilon}-{{
  \dot\epsilon}}^{2}+{ \ddot\epsilon}\,{\epsilon}}{{2{\epsilon}}^{2}}} 
\end{eqnarray}

\end{document}